\documentstyle[12pt,fleqn,epsfig]{article}
\begin{document}
\centerline{\bf{Ordinary muon capture as a probe of virtual transitions
of $\beta\beta$ decay}}
\bigskip
\centerline{Markus Kortelainen and Jouni Suhonen}
\medskip

\centerline{{\it{Department of Physics, University of Jyv\"askyl\"a}}}
\centerline{{\it{P.O.Box 35, FIN-40351, Jyv\"askyl\"a, Finland}}}
\bigskip
\bigskip

\noindent
{\bf Abstract}: {\it A reliable theoretical description of 
double-beta-decay processes needs
a possibility to test the involved virtual transitions
against experimental data. Unfortunately, only the lowest virtual 
transition can be probed by the traditional electron-capture or 
$\beta^-$-decay experiments. In this article we propose that 
calculated amplitudes for many virtual transitions
can be probed by experiments measuring rates of ordinary muon 
capture (OMC) to the relevant intermediate states. The first
results from such experiments are expected to appear soon.
As an example
we discuss the $\beta\beta$ decays of $^{76}$Ge and $^{106}$Cd
and the corresponding OMC for the $^{76}$Se and $^{106}$Cd
nuclei in the framework of the proton-neutron QRPA with realistic
interactions. It is found that the OMC observables, just like 
the $2\nu\beta\beta$-decay amplitudes, strongly depend on the 
strength of the particle-particle part of the proton-neutron 
interaction. }\\



The two-neutrino double beta ($2\nu\beta\beta$) and neutrinoless 
double beta ($0\nu\beta\beta$) decays proceed via virtual
transitions through states of the intermediate double-odd
nucleus. In the case of the $2\nu\beta\beta$ decay only the
intermediate $1^+$ states are involved, whereas in the case of
the $0\nu\beta\beta$ decay all the multipole states $J^{\pi}$
of the intermediate nucleus are active (see e.g. \cite{REPORT}).
Theoretical calculation of these virtual transitions has been
a hot subject since two decades and a host of 
different approaches have been devised to accomplish the task
\cite{REPORT}. The most relevant of these models, when only the
ground-state-to-ground-state $\beta\beta$ decays are considered, 
are the nuclear shell model and the proton-neutron random-phase
approximation (pnQRPA).

The calculation of the virtual $\beta$-decay type transitions is
a formidable task and experimental data on the corresponding
electron-capture (EC) or $\beta^-$-decay transitions from the
intermediate states to the initial or final ground state can help
in fine-tuning the model parameters before calculation of the
double-beta-decay rates. Unfortunately, the EC/$\beta^-$-type
of measurements can only probe the virtual transition through
the lowest intermediate $J^{\pi}$ state. A further experimental
analysis could be done by using the (p,n) or (n,p) charge-exchange
reactions \cite{EJI00} but the extraction of the relevant 
information is not streightforward.

In this Letter we propose to use the ordinary muon capture 
(OMC) as a probing tool for the nuclear wave functions involved
in the amplitudes of the virtual transitions of the $\beta\beta$
decay. The OMC process is like an EC process, except that the
mass of the captured negative muon, $\mu^-$, is about 200 times the
electron rest mass \cite{PRI59}. 
The relevant OMC transitions here are the ones
which start from the double-even $0^+$ ground state and end on
$J^{\pi}$ states of the intermediate double-odd nucleus. This means
that in the case of the $\beta^-\beta^-$ decays one probes the final
leg of a double-beta transition and in the case of the double
positron decays one probes the initial leg of the double-beta transition.
Due to the large mass energy absorbed by the final nucleus in an OMC 
transition, one is able to probe intermediate states at high excitation
energies, unlike in the EC transitions. A drawback of the OMC method,
from the theoretical point of view, is that in addition to the V-A
part the muon recoil activates also the induced parts 
(weak-magnetism and pseudoscalar terms) of the nucleonic current.
This means that the theoretical expression for the partial OMC rates
(i.e. capture rates to a particular final state)
is more complicated than the expressions for the virtual legs of
the $\beta\beta$ decay, and the OMC results as such can not be 
directly interpreted as amplitudes of the virtual transitions in
the $\beta\beta$ decay. Instead, the OMC rates can be used to test
in a versatile way the many-body wave functions of the $J^{\pi}$ 
states of the intermediate nucleus.

In the present work we demonstrate how one can use the partial OMC rates
to probe the structure of the $J^{\pi}$ states of the intermediate
nucleus. This is done by making calculations of the OMC rates and
the corresponding beta-decay rates in the framework of the proton-neutron
quasiparticle random-phase approximation (pnQRPA). 
As test cases we have chosen the final nucleus $^{76}$Se of the
$\beta^-\beta^-$ decay of $^{76}$Ge and the double electron-capture (ECEC)
decaying nucleus $^{106}$Cd. In the former case one can test by the OMC on
$^{76}$Se the states of $^{76}$As (this is the final virtual leg in
the $\beta^-\beta^-$ decay of $^{76}$Ge), the most relevant of which are
the $1^+$ states and the $2^-$ states (the ground state of $^{76}$As
is a $2^-$ state). In the latter case one can test the $1^+$ states
of $^{106}$Ag by the OMC on $^{106}$Cd (this is the initial virtual
leg in the ECEC decay of $^{106}$Cd). Also OMC rates to other
multipole states can be measured, but the above mentioned states ($1^+$
and $2^-$) have been chosen to serve as demonstration of the method.
In addition, some beta-decay data are available for the lowest
intermediate excitations in these nuclei so that one can probe to some
extent the quality of the calculations for the selected multipoles.

The experimental possibilities to use the OMC as a probe for the
virtual transitions of the $\beta\beta$ decays lie on prospects
to use enriched targets of nuclei like $^{48}$Ti, $^{76}$Se,
$^{82}$Kr, $^{96}$Mo, $^{100}$Ru, $^{106}$Cd, $^{116}$Sn, $^{130}$Xe,
etc. The need for an enriched ($A,N,Z$) target comes from the fact that
even a small admixture of the ($A+1,N+1,Z$) isotope would be very
dangerous, as the probability of the neutron-emission reaction

$\mu + (A+1,N+1,Z)\to \nu_{\mu} + (A+1,N+2,Z-1)\to \nu_{\mu} + 
(A,N+1,Z-1) + {\rm neutron}$

\noindent
is higher by an order of magnitude, and this channel would produce the
same excited states of the final $(A,N+1,Z-1)$ nucleus as the OMC in the
$(A,N,Z)$ nucleus. If the enrichment is less than $80\thinspace\%$ the
measurements should be done twice, for the $A$ and $A+1$ isotopes
separately, in order to separate the neutron-emission contamination.

Also the analysis of the gamma cascades following the capture
to an excited state in the double-odd daughter nucleus has to be
performed carefully. This kind of studies have been performed for the
1s-0d shell nuclei in T. P. Gorringe et al. \cite{GOR99} by measuring
38 gamma-ray lines and 29 ($\mu$,$\nu$) transitions following negative
muon capture on $^{24}$Mg, $^{28}$Si, $^{31}$P and $^{32}$S. In the
heavier, double-beta-decaying nuclei the density of the final states
of the muon capture is larger but there already exists a host of data 
on the energy levels and their gamma feeding in the intermediate
double-odd nuclei. It has to be noted that {\it not all} of the gamma
rays need to be detected. A set of the lowest ones already yields
information on the muon-capture rates to the lowest intermediate states
and can thus be used to probe the $\beta\beta$ virtual transitions 
to these states. The first experimental proposal concerning the 
measurement of the partial muon-capture rates in several double-beta-decay
camdidates, with mass numbers ranging from $A=48$ to $A=150$, has been
submitted to the Paul Scherrer Institut (PSI) in Villigen, Switzerland
by the Dubna-Louvain-Orsay-PSI-Jyv\"askyl\"a collaboration \cite{Slava}.

The $2\nu\beta\beta$ decay proceeds via the $1^+$ states of the
intermediate double-odd nucleus and the corresponding
expression for the inverse half-life can be written as

\begin{equation}
{[t_{1/2}^{(2 \nu)}(0_i^+ \rightarrow 0_f^+)]}^{-1}=
G_{\rm{DGT}}^{(2 \nu)} \; \;
{\mid M_{\rm {DGT}}^{(2 \nu)}\mid}^2 \; \; ,
\end{equation}
where $ G_{\rm{DGT}}^{(2 \nu)}$  is the integral
over the phase space of the leptonic variables \cite{REPORT,DOI85}.
The nuclear matrix element
$M_{\rm {DGT}}^{(2 \nu)}$ can be written as

\begin{equation} \label{eq:mdgt}
M_{ \rm {DGT}}^{(2 \nu)}= { { \sum_{m,n}
(0_f^{+} \mid \mid \sum_j \sigma(j) \tau^{\mp}_j
 \mid \mid 1_m^{+}) <1_m^+ \mid 1_n^+>
(1_n^{+} \mid \mid \sum_j \sigma(j) \tau^{\mp}_j
 \mid \mid 0_i^+) } \over {  ( {{1} \over {2}} Q_{\beta \beta}+
E_m -M_i)/ m_{\rm e} +1  }}
   \;  \;  ,
\end{equation}
where the transition operators are the usual Gamow-Teller operators,
$Q_{\beta \beta}$ is the $2\nu\beta\beta$ $Q$ value, $E_m$ is the
energy of the $m$th intermediate state, $M_i$ is the mass energy
of the initial nucleus, and $m_{\rm e}$ is the electron-mass energy.
The overlap $<1_m^+ \mid 1_n^+>$
between the two sets of $1^+$ states, which are pn-QRPA solutions
based on the initial and final ground states, helps in matching
the two branches of virtual excitations \cite{TOM91}.

The expressions for the $0\nu\beta\beta$ amplitudes are more involved
and are given e.g. in \cite{REPORT,DOI85,TOM91}. In this case all
the intermediate multipoles are involved in the calculation
of the half-life and the transition operators in the virtual 
amplitudes contain spherical harmonics coming from the multipole expansion
of the neutrino-exchange potential.

The formalism needed for the calculation of the OMC rate
is developed in Ref. \cite{MOR60}. Here we cite the main results
of Ref. \cite{MOR60}, according to which
we can write an explicit formula for the capture rate as
\begin{equation} \label{eq:transw}
W = 4P(\alpha Z m_{\mu}')^{3}\frac{2J_{\rm f}+1}{2J_{\rm i}+1}
\left(1-\frac{q}{m_{\mu} +AM} \right)q^{2},
\end{equation}
with $A$ being the mass number of the initial and final nuclei, $Z$
the charge number of the initial nucleus, and
$m_{\mu}'$ is the reduced muon mass. Furthermore, $\alpha$ denotes the
fine-structure constant, $M$ the average nucleon mass, $m_{\mu}$ the
muon mass, and $q$ the magnitude of the exchanged momentum between the
captured muon and the nucleus.
The term $P$ in equation (\ref{eq:transw}) can be written as
\begin{eqnarray} \label{eq:muupee}  
P &=& \sum_{\kappa u}| g_{\rm{V}} {\mathcal M}[0lu]S_{0u}(\kappa) \delta_{lu} \nonumber \\
 & & -g_{\rm{A}}{\mathcal M}[1lu]S_{1u}(\kappa) - \frac{g_{\rm{V}}}{M}
      {\mathcal M}[1\bar{l}up]S_{1u}'(-\kappa) \nonumber \\
 & & + \sqrt{3}(g_{\rm{V}}Q/2M) \left( \sqrt{(\bar{l}+1)/(2\bar{l}+3)}
      {\mathcal M}[0\bar{l}+1 u+] \delta_{\bar{l}+1,u} \right.\nonumber \\
 & & +\left. \sqrt{\bar{l}/(2\bar{l}-1)}{\mathcal M}[0\bar{l}-1 u-]
     \delta_{\bar{l}-1,u}  \right) S_{1u}'(-\kappa) \nonumber \\
 & & + \sqrt{\frac{3}{2}}(g_{\rm{V}}Q/M)(1+\mu_{\rm{p}}-\mu_{\rm{n}}) \left(
     \sqrt{\bar{l}+1}W(11u\bar{l},1\bar{l}+1){\mathcal M}[1\bar{l}+1u+] \right. \nonumber \\
 & & +\left. \sqrt{\bar{l}}W(11u\bar{l},1,\bar{l}-1) {\mathcal M}
     [1,\bar{l}-1u-] \right) S_{1u}'(-\kappa) \nonumber \\
 & & +(g_{\rm{A}}/M){\mathcal M}[0\bar{l}up]S_{0u}'(-\kappa) \delta_{\bar{l}u}
     + \sqrt{\frac{1}{2}}(g_{\rm{A}}-g_{\rm{P}})(Q/2M) \nonumber \\
 & & \times \left( \sqrt{(\bar{l}+1)/(2\bar{l}+1)} {\mathcal M}[1\bar{l}+1u+]
     + \sqrt{\bar{l}/(2\bar{l}+1)}{\mathcal M}[1\bar{l}-1u-] \right) \nonumber \\
 & & \times S_{0u}'(-\kappa)\delta_{\bar{l}u} |^{2}\ ,
\end{eqnarray}
where the $W()$ symbols are the usual Racah coefficients, and the
definition for $\bar{l}$, the matrix elements ${\mathcal M}[kwu($$\pm \atop p$$)]$ 
and the geometrical factors $S_{ku}(\kappa )$ and $S'_{ku}(-\kappa )$ can be 
found in Refs. \cite{MOR60,SII98}. 
      
Finally, we note that a rather direct bridge between the $\beta\beta$ and 
the OMC processes can be established in the limit $q,Z \to 0$, 
since from the definition of the OMC matrix 
elements in Ref. \cite{MOR60,SII98} it follows that 
\begin{equation}
[101] \to \hat{J_{\rm f}}\frac{\sqrt{3}}{4 \pi}M_{\rm GT},
\end{equation}
where $M_{\rm GT}$ is the reduced matrix element 
\begin{equation}
M_{\rm GT} = (1^{+}_n|| \sum_{j} \sigma(j) \tau^{+}_{j} ||0^{+})
\end{equation}
of Eq. (\ref{eq:mdgt}). Here the $0^+$ state denotes either the final
(for $\beta^-\beta^-$ decay) or initial 
(for $\beta^+/{\rm EC}\beta^+/{\rm EC}$ decay)
$0^+$ state of the $\beta\beta$-decay process.

In the numerical computations we have used the proton model space 
1p-0f-2s-1d-0g-0h$_{11/2}$ both for the $A=76$ isobars and the
$A=106$ isobars. On the neutron side the same model space was used
for the $A=76$ isobars, but for the $A=106$ isobars the extended
neutron space 1p-0f-2s-1d-0g-2p-1f-0h was adopted. The corresponding
single-particle energies were obtained by using the Woods-Saxon
(WS) well with the parametrization of \cite{BOH69}. This basis
was used recently in Ref. \cite{SUH00} for calculation of
$0\nu\beta\beta$-decay rates in the $A=76$ system. For the $A=106$
system, the neutron and proton WS single-particle energies were
adjusted near the corresponding Fermi surfaces according to data on 
neighboring proton-odd and neutron-odd nuclei. This is the same basis
which was used in \cite{SUH01} for a succesful description of the
beta and double-beta data on the $A=106$ isobars. 

The nuclear hamiltonian of the two isobaric regions was obtained from
the Bonn one-boson-exchange potential \cite{HOL81} with 
empirical renormalization by using the phenomenological pairing
gaps, giant Gamow-Teller resonances, and spectroscopic data on
nuclei close to the relevant isobars. For more details the reader is
referred to the articles \cite{SUH00,SUH01}. As in very many
calculations in the recent past \cite{REPORT} also the results of the 
present pnQRPA calculation in the $1^+$ channel for the first intermediate
$1^+$ state depend strongly on the strength parameter $g_{\rm pp}$ of
the particle-particle part of the proton-neutron interaction. This
strong dependence on $g_{\rm pp}$ also strongly influences the 
$2\nu\beta\beta$-decay rates leading to a strong suppression of the
$2\nu\beta\beta$ matrix element, as first discussed in \cite{GPP}. For
that reason, to demostrate the corresponding effects on the OMC
observables, we present our analysis as a function of $g_{\rm pp}$
for the $A=76$ isobars. For the $A=106$ isobars we adopt the value
$g_{\rm pp}=0.8$ coming from the exhaustive spectroscopy analysis of
Ref. \cite{SUH01}.

\begin{figure}[h]
\mbox{\epsfig{file=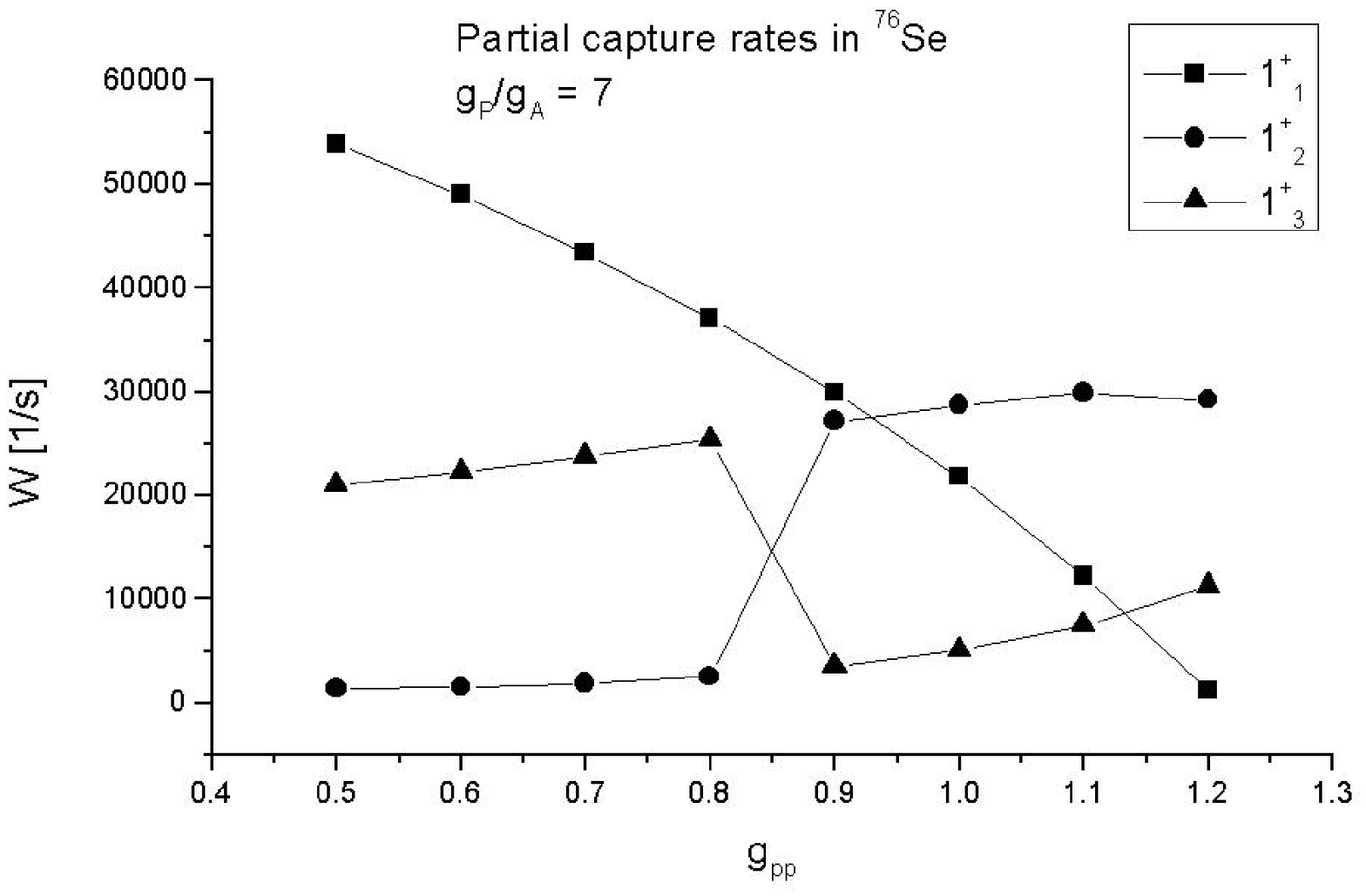,width=10cm}}
\caption{Partial capture rates $^{76}$Se($0^{+}_{\rm g.s.}$)
$\to$ $^{76}$As($1^{+}_{i}$) as functions of $g_{\rm pp}$ with
$g_{\rm P}/g_{\rm A}$ = 7. }
\label{f:se1p}
\end{figure}

\begin{figure}[h]
\mbox{\epsfig{file=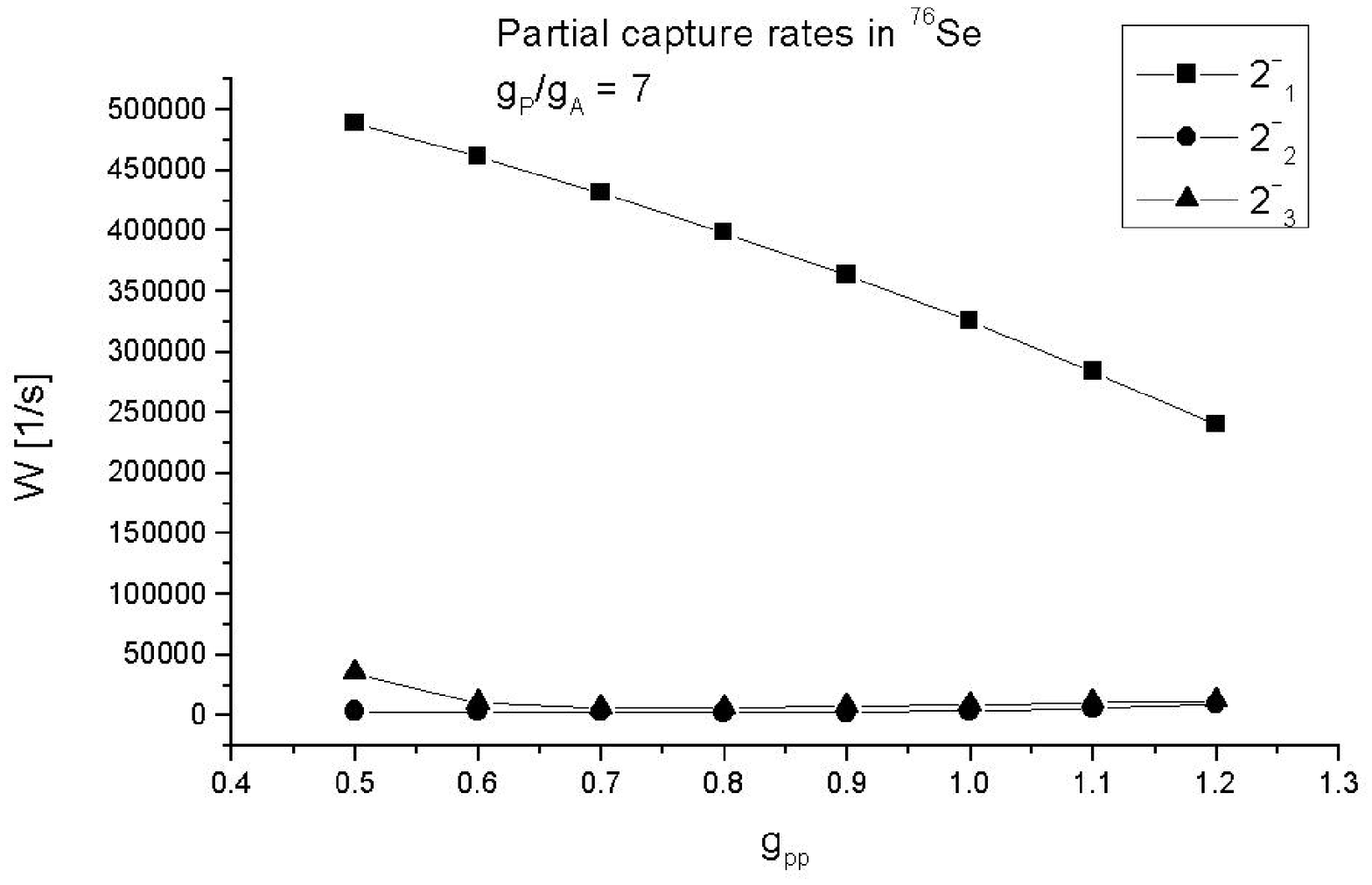,width=10cm}}
\caption{Partial capture rates $^{76}$Se($0^{+}_{\rm g.s.}$)
$\to$ $^{76}$As($2^{-}_{i}$) as functions of $g_{\rm pp}$ with
$g_{\rm P}/g_{\rm A}$ = 7.}
\label{f:se2m}
\end{figure}

\begin{figure}[h]
\mbox{\epsfig{file=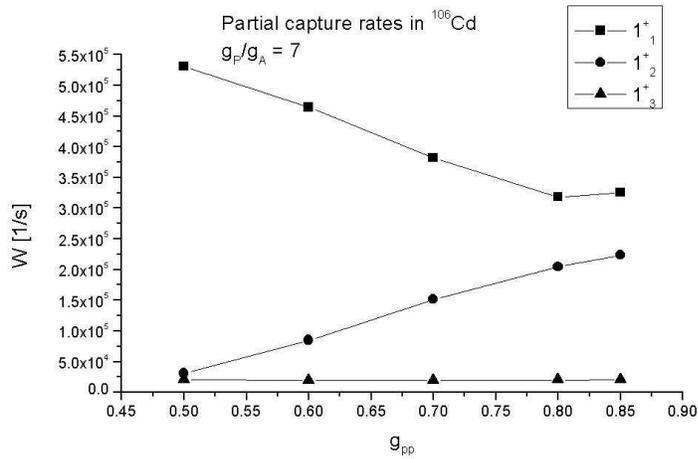,width=10cm}}
\caption{Partial capture rates $^{106}$Cd($0^{+}_{\rm g.s.}$)
$\to$ $^{106}$Ag($1^{+}_{i}$) as functions of $g_{\rm pp}$ with
$g_{\rm P}/g_{\rm A}$ = 7.}
\label{f:cd1p}
\end{figure}

Partial muon-capture rates $W$ were calculated 
for transitions $^{76}$Se($0^{+}_{\rm g.s.}$)
$\to$ $^{76}$As($1^{+},2^{-}$) and $^{106}$Cd($0^{+}_{\rm g.s.}$) $\to$ 
$^{106}$Ag($1^{+}$). In these calculations we have used the 
Goldberger-Treiman (PCAC) value $g_{\rm P}/g_{\rm A} = 7$
for the ratio of the coupling strengths of the induced pseudoscalar and
axial-vector parts of the charged weak current. More discussion of this
matter can be found in \cite{SII98}. Furthermore, the nuclear-matter
renormalized value of $g_{\rm A} = 1.0$ was adopted for the axial-vector
strength. We have plotted the correponding results in Figs. 
\ref{f:se1p}--\ref{f:cd1p} as a function of $g_{\rm pp}$. Here one can clearly
see that the capture rates to the $1^+_1$ and $2^-_1$ states are strongly
dependent on the value of the particle-particle strength $g_{\rm pp}$ of
the proton-neutron force. This dependence for the higher-lying states is
smaller, although a notable exception is the OMC to the $1^+_2$ state in
$^{106}$Ag, as seen in Fig. \ref{f:cd1p}. In Fig. \ref{f:se1p} one can
see that the states $1^+_2$ and $1^+_3$ cross at about $g_{\rm pp}=0.85$,
and that for $g_{\rm pp}$ values beyond 1.1 one obtains very small values
for the OMC rate to the first $1^+$ state. Beyond this point
the recoil matrix element $[011p]$ grows unphysically large and cancels
against the contributions coming from the other nuclear matrix elements.
The unphysically large value of the recoil matrix element stems from 
the excessive growth of the ground-state correlations, leading to the
same order of magnitude for the forward- and backward-going amplitudes
in the pnQRPA approach. This happens close to the breaking point of the
pnQRPA, which in this case is around the value $g_{\rm pp}\simeq 1.2$.

Partial capture rates as well as
the corresponding $\log ft$ values for $\beta^{-}$ transitions, for specific values
of $g_{\rm pp}$, can be found in Table \ref{t:logftw}. The 
value $g_{\rm pp}=0.9$ for the $1^+$ channel in $A=76$ was chosen because
it would yield a sensible log$ft$ value for the $\beta^-$ decay of the
$1^+_1$ state (for $g_{\rm pp}=1.0$ the log$ft$ is far too large). The 
log$ft$ value for the unique first-forbidden $\beta^-$ transition from 
the $2^-_1$ state is practically independent of the value of $g_{\rm pp}$
and thus the unrenormalized value of $g_{\rm pp}=1.0$ was chosen. For
the $A=106$ isobars the value $g_{\rm pp}=0.8$ is based on spectroscopy
\cite{SUH01}, as discussed earlier. The obtained log$ft$ values can be
compared with the available data, namely
$\log ft(^{76}$As$(2^{-}_{1}) \to ^{76}$Se$(0^{+}_{\rm g.s.}))  =  9.7 $ 
and $\log ft(^{106}$Ag$(1^{+}_{1}) \to  ^{106}$Cd$(0^{+}_{\rm g.s.})) >  4.2 $
\cite{FIRESTONE}. Thus, a reasonable agreement with available data is
obtained, and the predicted OMC rates can be used for estimating the
possibility of experimental determination of the discussed partial OMC rates.
In addition to the quoted partial OMC rates, we list in Table \ref{t:totalW}
the total muon-capture rates for each calculated $J^{\pi}$ multipole.
The evaluation  of the total rates has been done for the above discussed
values of $g_{\rm pp}$. The total OMC rates, like the partial ones, are 
dependent on the value of $g_{\rm pp}$.
 
\begin{figure}[h]
\mbox{\epsfig{file=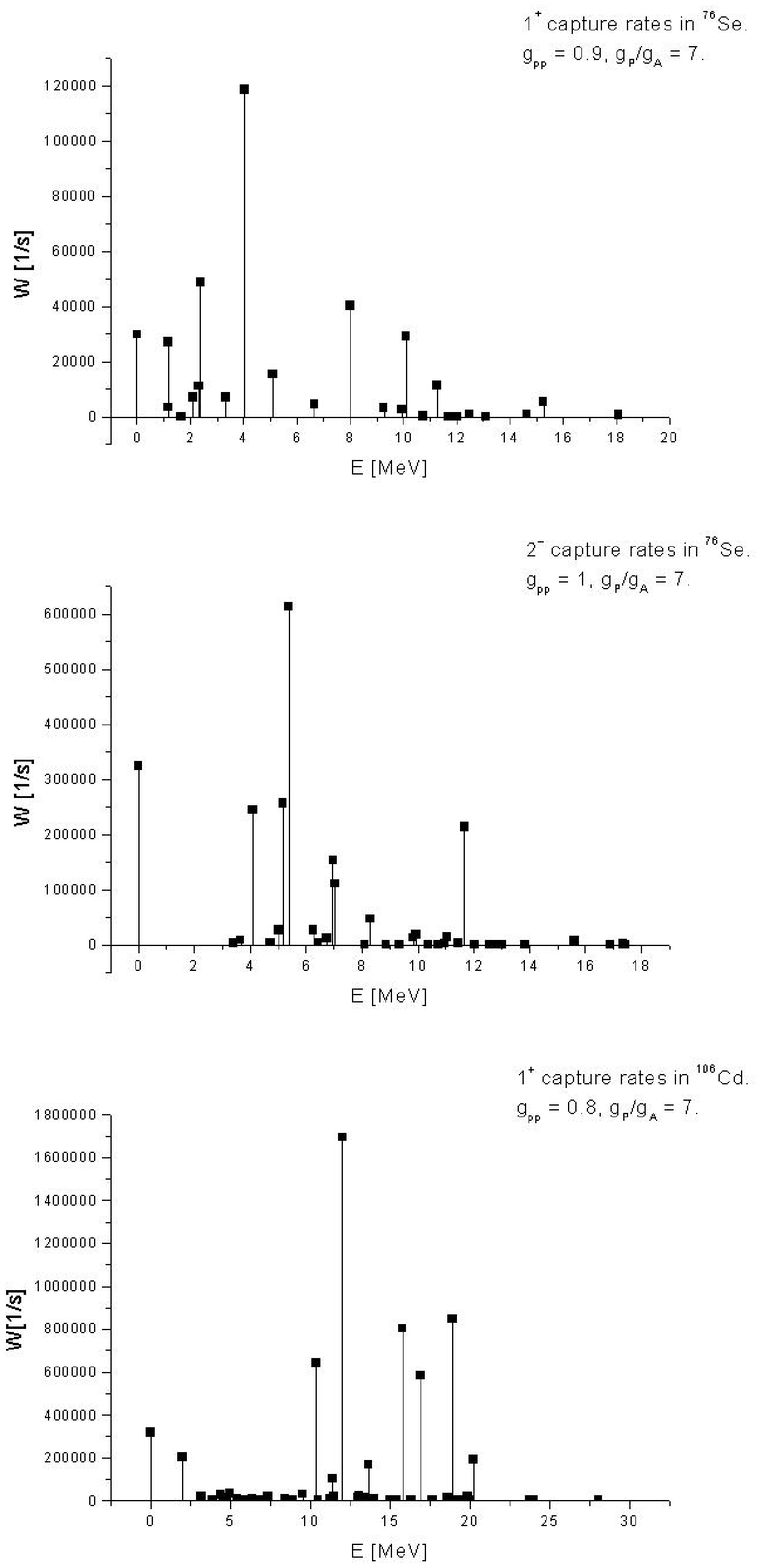,width=10cm}}
\caption{Partial capture rates as functions of the excitation energy 
in the final nucleus.}
\label{f:tolpat}
\end{figure}

In Fig. \ref{f:tolpat} we present histogams of all calculated partial 
capture rates with the correspondig excitation energy in the final nucleus.
The total capture rates of Table \ref{t:totalW} have been obtained by
integrating these distributions over the energy. As can be seen, the 
OMC rate distribution for $1^+$ capture is very different for the
$A=76$ and the $A=106$ isobars, the distribution peaking strongly
between $10\thinspace {\rm MeV}$ and $20\thinspace {\rm MeV}$ for the
$A=106$ system and below $4\thinspace {\rm MeV}$ for the $A=76$ system.
The $2^-$ distribution concentrates on low energies, the $2^-_1$
contribution being notable.

In conclusion, in the present work we have studied the OMC rates as
tests of the nuclear wave functions involved in the amplitudes of
the virtual transitions of the two-neutrino and neutrinoless double 
beta decay. It was found that the OMC observables 
(partial and total capture rates
to $1^+$ and $2^-$ states) depend on the strength ($g_{\rm pp}$) of the
particle-particle part of the proton-neutron interaction. 
The same was observed for the $2\nu\beta\beta$-decay amplitudes already
more than a decade ago. The
strongest dependence is associated with the decays via the first $1^+$
state, whereas the dependence of the $2^-$ states on $g_{\rm pp}$
is much weaker than for the $1^+$ states. On the experimental side,
the first results from the measurements of the partial muon-capture
rates to the double-odd intermediate nuclei involved in the virtual
transitions of the double beta decay will be available soon. The
first experimental proposal has been submitted to the PSI, including the
$A=76$ and $A=106$ systems discussed in this article. \\

{\bf Acknowledgements} 

The authors thank Dr. V. Egorov for useful discussions on the 
experimental aspects of this work.
This work has been supported by the Academy of Finland under the
Finnish Centre of Excellence Programme 2000-2005 (Project No.
44875, Nuclear and Condensed Matter Programme at JYFL). \\

\newpage

\centerline{\bf Tables}

\bigskip

\begin{table}[h]
\caption{$\log ft$ values of $\beta^{-}$ decays of three lowest states
of the $1^+$ and $2^-$ multipolarities, together with the corresponding
muon capture rates $W$. For $A=76$ the values
$g_{\rm pp}$ = 0.9 ($J^{\pi}=1^{+}$) and $g_{\rm pp}$ = 1.0 
($J^{\pi}=2^{-}$) were used. For $A=106$ the value $g_{\rm pp}$ = 0.8
was adopted. The ratio $g_{\rm P}/g_{\rm A}$ = 7 was assumed
in the calculations. }
\label{t:logftw}
\begin{tabular}{|l|c|c|c|c|c|c|c|c|c|}
\hline
& \multicolumn{6}{|c|}{$A=76$} & \multicolumn{3}{|c|}{$A=106$} \\
Quantity & $1^{+}_{1}$ & $1^{+}_{2}$ & $1^{+}_{3}$ & $2^{-}_{1}$ & $2^{-}_{2}$ & 
   $2^{-}_{3}$ & $1^{+}_{1}$ & $1^{+}_{2}$ & $1^{+}_{3}$ \\ \hline
$\log ft(J_{\rm i}^{\pi} \to 0^{+}_{\rm g.s.})$ & 5.77 & 6.73 & 7.25 &
   8.73 & 9.25 & 9.26 & 4.30 & 4.08 & 5.83 \\ \hline
$W$ [$10^{3}$ 1/s] & 31.17 & 23.79 & 1.92 & 301.92 & 2.13 & 9.08 & 325.52 & 
   211.71 & 19.04 \\
\hline
\end{tabular}
\end{table}

\begin{table}[h]
\caption{Total muon capture rates with $g_{\rm P}/g_{\rm A}$ = 7 for the
$1^+$ and $2^-$ multipoles.}
\label{t:totalW}
\begin{tabular}{|r|c|l|}
\hline
Captures & $W_{\rm tot}$ [$10^{3}$ 1/s] & $g_{\rm pp}$ \\ 
\hline
$^{76}$Se($0^{+}_{\rm g.s.}$) $\to$ $^{76}$As($1^{+}$) & 359 & 0.9 \\ \hline
$^{76}$Se($0^{+}_{\rm g.s.}$) $\to$ $^{76}$As($2^{-}$) & 2020 & 1.0 \\ \hline
$^{106}$Cd($0^{+}_{\rm g.s.}$) $\to$ $^{106}$Ag($1^{+}$) & 5385 & 0.8 \\ \hline
\end{tabular}
\end{table}

\newpage

\newpage

\centerline{\bf Figures}

\bigskip

\end{document}